\noindent 
\centerline{\bf Functions of Mittag-Leffler and Fox:}
\centerline{\bf The Pathway Model to Tsallis Statistics and Beck-Cohen Superstatistics}

\vskip.3cm 
\centerline{A.M. Mathai} 
\vskip.0cm
\centerline{Centre for Mathematical Sciences Pala Campus}
\vskip.0cm
\centerline{Arunapuram P.O., Pala, Kerala-686574, India}
\vskip.0cm
\centerline{and Department of Mathematics and Statistics}
\vskip.0cm
\centerline{McGill University, Canada}
\vskip.0cm
\centerline{mathai@math.mcgill.ca; cmspala@gmail.com}

\vskip.3cm 
\noindent 
\centerline{and}

\vskip.2cm 
\noindent 
\centerline{H.J. Haubold}
\vskip.0cm
\centerline{Office for Outer Space Affairs, United Nations}
\vskip.0cm 
\centerline{Vienna International Centre, P.O. Box 500,}
\vskip.0cm
\centerline{A-1400 Vienna, Austria}
\vskip.0cm
\centerline{and Centre for Mathematical Sciences Pala Campus}
\vskip.0cm
\centerline{Arunapuram P.O., Pala, Kerala-686574, India}
\vskip.0cm
\centerline{hans.haubold@unvienna.org; cmspala@gmail.com}

\vskip.5cm 
\noindent Keywords: generalized hypergeometric functions, Mathai's pathway
model, Tsallis statistics, Beck-Cohen superstatistics, Mellin-Barnes integrals,
fractional integrals, fractional differential equations.

\vskip.3cm 
\noindent AMS Subject classification: 33C60, 82C31, 62E15

\vskip.3cm 
\noindent {\bf Abstract}

\vskip.3cm In reaction rate theory, in production-destruction type models and
in reaction-diffusion problems when the total derivatives are
replaced by fractional derivatives the solutions are obtained in
terms of Mittag-Leffler functions and their generalizations. When
fractional calculus enters into the picture the solutions of these
problems, usually available in terms of generalized hypergeometric functions, 
switch to Mittag-Leffler functions and their
generalizations into Wright functions and subsequently into Fox functions. In this paper, connections
are established among generalized Mittag-Leffler functions, Mathai's pathway
model, Tsallis statistics, Beck-Cohen superstatistics, and among corresponding 
entropic measures. The Mittag-Leffler function, for large values of the 
parameter, approaches a power-law. For values of the parameter close to 
zero, the Mittag-Leffler function behaves like a stretched exponential. 
The Mittag-Leffler function is a generalization of the exponential function 
and represents a deviation from the exponential paradigm whenever it shows 
up in solution of physical problems. The paper elucidates the relation between 
analytic representations of the q-exponential function that is fundamental 
to Tsallis statistics, Mittag-Leffler, Wright, and Fox functions, respectively, 
utilizing Mellin-Barnes integral representations. 

\vskip.3cm \noindent {\bf 1.\hskip.3cm Introduction}

\vskip.3cm Fundamental laws of physics are written as equations for
the time evolution of a quantity $x(t)$,

$${{{\rm d}x(t)}\over{{\rm d}t}}=Ax(t),\eqno(1.1)
$$where if $A$ is  limited to a linear operator we have Maxwell's
equation or Schroedinger equation, or it could be Newton's law of
motion or Einstein's equations for geodesics if $A$ may also be a
nonlinear operator [8]. When $A$ is linear then the mathematical
solution is

$$x(t)=x_0{\rm e}^{-At}\eqno(1.2)
$$where $x_0$ is the initial value at $t=0$. Taking the simplest case of (1.1), as used in reaction rate
theory ([8],[4]), if the number density at time $t$ of the $i$-th particle is
$N_i(t)$ and if the number of particles produced or the production
rate is proportional to $N_i(t)$ then the reaction equation is

$${{{\rm d}N(t)}\over{{\rm d}t}}=k_1N(t), ~k_1>0
$$deleting $i$ for convenience. If the decay rate is also
proportional to $N(t)$ then the corresponding equation is

$${{{\rm d}N(t)}\over{{\rm d}t}}=-k_2N(t), ~k_2>0.
$$Then the residual effect in a production-destruction mechanism is
of the form

$${{{\rm d}N(t)}\over{{\rm d}t}}=-c N(t), c>0\Rightarrow
N(t)-N_0=-c\int N(t){\rm d}t\eqno(1.3)
$$if the destruction rate dominates so that the production-destruction model
is a decaying model.  Such production-destruction models abound in scientific
disciplines. If the total integral or the total derivative in (1.3)
is replaced by a fractional integral then we have

$$N(t)-N_0=-c^{\nu}~{_0D}_t^{-\nu}N(t),\eqno(1.4)
$$where ${_0D}_t^{-\nu}$ is the Riemann-Liouville fractional
integral operator ([5],[7])defined by

$${_0D}_t^{-\nu}f(t)={{1}\over{\Gamma(\nu)}}\int_0^t(t-u)^{\nu-1}f(u){\rm
d}u\eqno(1.5)
$$for $\Re(\nu)>0$ where $\Re(\cdot)$ denotes the real part of
$(\cdot)$, $c$ is replaced by $c^{\nu}$ for convenience, and then
the solution of (1.4) goes into the category of a Mittag-Leffler
function [4], namely,

$$N(t)=N_0\sum_{k=0}^{\infty}{{(-c^kt^k)^{\nu}}\over{\Gamma(1+k\nu)}}=N_0~E_{\nu}[-(ct)^{\nu}]\eqno(1.6)
$$where $E_{\nu}(\cdot)$ is the Mittag-Leffler function. The
generalized Mittag-Leffler function is defined as

$$E_{\alpha,\beta}^{\gamma}(z)=\sum_{k=0}^{\infty}{{(\gamma)_k}\over{k!\Gamma(\beta+\alpha
k)}},~\Re(\alpha)>0, \Re(\beta)>0,\eqno(1.7)
$$and when $\Gamma(\gamma)$ is defined, it has the following
Mellin-Barnes representation

$$\eqalignno{E_{\alpha,\beta}^{\gamma}(z)&={{1}\over{\Gamma(\gamma)}}\sum_{k=0}^{\infty}{{\Gamma(\gamma+k)}\over{k!\Gamma(\beta+\alpha
k)}}&(1.8)\cr &={{1}\over{\Gamma(\gamma)}}{{1}\over{2\pi
i}}\int_{c-i\infty}^{c+i\infty}{{\Gamma(s)\Gamma(\gamma-s)}\over{\Gamma(\beta-\alpha
s)}}(-z)^{-s}{\rm d}s&(1.9)\cr}
$$for $0<c<\Re(\gamma), \Re(\gamma)>0, i=\sqrt{-1}$. Some special
cases of the generalized Mittag-Leffler function are the following:

$$\eqalignno{E_{\alpha,\beta}^1(z)&=E_{\alpha,\beta}(z)=\sum_{k=0}^{\infty}{{1}\over{\Gamma(\beta+k\alpha)}}z^k&(1.10)\cr
E_{\alpha,1}^1&=E_{\alpha,1}(z)=E_{\alpha}(z)=\sum_{k=0}^{\infty}{{z^k}\over{\Gamma(1+k\alpha)}}&(1.11)\cr}
$$and when $\alpha=1$ we have

$$E_{\alpha}(z)=E_1(z)={\rm e}^z.\eqno(1.12)
$$Thus the Mittag-Leffler function can be looked upon as an
extension of the exponential function. The Mellin-Barnes
representation in (1.9) is a special case of the Mellin-Barnes
representation of the Wright's function [3] which is defined
as

$$\eqalignno{{_p\psi_q}(z)&={_p\psi_q}\left[z\bigg\vert_{(b_j,\beta_j),j=1,...,q}^{(a_j,\alpha_j),j=1,...,p}\right]\cr
&=\sum_{k=0}^{\infty}{{\prod_{j=1}^p\Gamma(a_j+\alpha_j
k)}\over{\prod_{j=1}^q\Gamma(b_j+\beta_j
k)}}{{z^k}\over{k!}}&(1.13)\cr &={{1}\over{2\pi
i}}\int_{c-i\infty}^{c+i\infty}\Gamma(s)\left\{{{\prod_{j=1}^p\Gamma(a_j-\alpha_j
s)}\over{\prod_{j=1}^q\Gamma(b_j-\beta_j s)}}\right\}(-z)^{-s}{\rm
d}s&(1.14)\cr}
$$where $0<c<\min_{1\le j\le p}\Re\left({{a_j}\over{\alpha_j}}\right)$
with $a_j,j=1,...,p$ and $b_j,j=1,...,q$ being complex quantities
and $\alpha_j>0,j=1,...,p$ and $\beta_j>0,j=1,...,q$ being real
quantities.  Observe from (1.14) that Wright's function is a special
case of the H-function [7] and the H-function is defined as
the following Mellin-Barnes integral

$$\eqalignno{H_{p,q}^{m,n}(z)&=H_{p,q}^{m,n}\left[z\bigg\vert_{(b_1,\beta_1),...,(b_q,\beta_q)}^{(a_1,\alpha_1),...,(a_p,\alpha_p)}\right]
={{1}\over{2\pi i}}\int_{c-i\infty}^{c+i\infty}\phi(s)z^{-s}{\rm
d}s&(1.15)\cr \noalign{\hbox{where}}
\phi(s)&={{\left\{\prod_{j=1}^m\Gamma(b_j+\beta_js)\right\}\left\{\prod_{j=1}^{n}\Gamma(1-a_j-\alpha_js)\right\}}
\over{\left\{\prod_{j=m+1}^q\Gamma(1-b_j-\beta_js)\right\}\left\{\prod_{j=n+1}^p\Gamma(a_j+\alpha_js)\right\}}},&(1.16)\cr}
$$for $\max_{1\le j\le m}{{\Re(-b_j)}\over{\beta_j}}<c<\min_{1\le
j\le n}{{\Re(1-a_j)}\over{\alpha_j}}$ where $a_j,j=1,...,p$ and
$b_j,j=1,...,q$ are complex quantities, $\alpha_j>0,j=1,...,p$ and
$\beta_j>0,j=1,...,q$ re real quantities. Existence conditions and
various contours may be seen from books on the H-function, for example,
([5],[7]). Equations (1.8), (1.14), and (1.15) establish the relations between Mittag-Leffler, Wright, and Fox functions 
to be used in the following for q-exponential function in Tsallis statistics, Beck-Cohen superstatistics, 
and their link to Mathai's pathway model.

\vskip.3cm \noindent \vskip.3cm \noindent {\bf 1.1.\hskip.3cm
Extension of the reaction rate model and Mittag-Leffler function}

\vskip.3cm The reaction rate model in (1.3) can be extended in
various directions ([8],[4]). For example, if $N_0$ is replaced by $N_0f(t)$
where $f(t)$ is a general integrable function on the finite interval
$[0,b]$ then it is easy to see that for the solution of the equation

$$N(t)-N_0f(t)=-c^{\nu}~{_0D}_t^{-\nu}N(t)\eqno(1.17)
$$there holds the formula

$$N(t)=cN_0\int_0^tH_{1,2}^{1,1}\left[c^{\nu}(t-x)^{\nu}\bigg\vert_{(-{{1}\over{\nu}},1),(0,\nu)}^{(-{{1}\over{\nu}},1)}\right]f(x){\rm
d}x.\eqno(1.18)
$$Some special cases are the following: Let $\nu>0,\rho>0,c>0$. Then
for the solution of the fractional equation

$$N(t)-N_0t^{\rho-1}=-c^{\nu}~{_0D}_t^{-\nu}N(t)\eqno(1.19)
$$there holds the formula

$$N(t)=N_0\Gamma(\rho)t^{\rho-1}E_{\nu,\rho}(-(ct)^{\nu}).\eqno(1.20)
$$Let $c>0,\nu>0,\mu>0$. Then for the solution of the fractional
equation

$$N(t)-N_0t^{\mu-1}E_{\nu,\mu}^{\gamma}[-(ct)^{\nu}]=-c^{\nu}~{_0D}_t^{-\nu}N(t)\eqno(1.21)
$$there holds the formula

$$N(t)=N_0t^{\mu-1}E_{\nu,\mu}^{\gamma+1}[-(ct)^{\nu}].\eqno(1.22)
$$

\vskip.3cm \noindent {\bf 1.2.\hskip.3cm Fractional partial
differential equations and Mittag-Leffler functions}

\vskip.3cm In a series of papers, see for example, ([2],[4])
it is illustrated that the solutions of certain fractional partial
differential equations, resulting from fractional diffusion
problems, are available in terms of Mittag-Leffler functions. For
example, consider the equation

$${_0D}_t^{\nu}N(x,t)-{{t^{-\nu}}\over{\Gamma(1-\nu)}}\delta(x)=-c^{\nu}{{\partial^2}\over{\partial
x^2}}N(x,t)\eqno(1.23)
$$with initial conditions

$${_0D}_t^{\nu-k}N(x,t)|_{t=0}=0,k=1,...,n
$$where $n=[\Re(\nu)]+1, c^{\nu}$ is the diffusion constant,
$\delta(x)$ is the Dirac's delta function and $[\Re(\nu)]$ is the
integer part of $\Re(\nu)$. The solution of (1.23), by taking
Laplace transform with respect to $t$ nd Fourier transform with
respect to $x$ and then inverting, can be shown to be of the form

$$N(x,t)={{1}\over{(4\pi
c^{nu}t^{\nu})^{1\over2}}}H_{1,2}^{2,0}\left[{{|x|^2}\over{4c^{\nu}t^{\nu}}}\bigg\vert_{(0,1),({1\over2},1)}^{(1-{{\nu}\over2},\nu)}\right]\eqno(1.24)
$$which in special cases reduce to Mittag-Leffler functions.

\vskip.2cm This paper is organized as follows: Section 2 gives the
classical special function technique of eliminating the upper or
lower parameters from a general hypergeometric series. Section 3
establishes the pathways of going from Mittag-Leffler function to
pathway models to Tsallis statistics and Beck-Cohen superstatistics through the
parameter elimination technique. Section 4 gives representations of
Mittag-Leffler functions and pathway model in terms of H-functions
and then gives a pathway to go from a Mittag-Leffler function to the
pathway model through H-function.

\vskip.3cm \noindent {\bf 2.\hskip.3cm A classical special function
technique}

\vskip.3cm The well established technique of eliminating a numerator or
denominator parameter from a general hypergeometric function in the
classical theory of special functions, is the following: For
convenience, we will illustrate it on a confluent hypergeometric
function.

$${_1F_1}(a;b;z)=\sum_{k=0}^{\infty}{{(a)_k}\over{(b)k}}{{z^k}\over{k!}},(a)_m=a(a+1)...(a+m-1),a\ne0,(a)_0=1.\eqno(2.1)
$$Observe that

$$\eqalignno{{{(a)_k}\over{a^k}}&={{a}\over{a}}{{(a+1)}\over{a}}...{{(a+m-1)}\over{a}}\cr
&=1(1+{{1}\over{a}})(1+{{2}\over{a}})...(1+{{m-1}\over{a}})\rightarrow
1\hbox{  as  }a\rightarrow\infty&(2.2)\cr \noalign{\hbox{for all
finite $k$. Similarly}} {{b^k}\over{(b)_k}}&\rightarrow 1\hbox{ when
}b\rightarrow\infty.&(2.3)\cr}
$$Therefore

$$\lim_{a\rightarrow\infty}{_1F_1}(a;b;{{z}\over{a}})={_0F_1}(~~;b;z)\eqno(2.4)
$$which is a Bessel function. Thus we can go from a confluent
hypergeometric function to a Bessel function through this process.
Similarly

$$\lim_{b\rightarrow\infty}{_1F_1}(a;b;bz)={_1F_0}(a;~~:z)=(1-z)^{-a},~|z|<1.\eqno(2.5)
$$Thus, from a confluent hypergeometric function we can go to a
binomial function. Further,

$$\eqalignno{\lim_{b\rightarrow\infty}{_0F_1}(~~;b;bz)&={_0F_0}(~~;~~;z)={\rm
e}^z&(2.6)\cr \noalign{\hbox{and}}
\lim_{a\rightarrow\infty}{_1F_0}(a;~~;{{z}\over
{a}})&={_0F_0}(~~;~~;z)={\rm e}^z.&(2.7)\cr}
$$Thus, we can go from a Bessel function as well as from a binomial
function to an exponential function. These two results can be stated
in a slightly different form as follows:

$$\eqalignno{\lim_{q\rightarrow
1}{_0F_1}(~~~;{{1}\over{q-1}};-{{z}\over{q-1}})&={\rm
e}^{-z}&(2.8)\cr \noalign{\hbox{and}} \lim_{q\rightarrow
1}{_1F_0}({{1}\over{q-1}};~~;-(q-1)z)&=\lim_{q\rightarrow
1}[1+(q-1)z]^{-{{1}\over{q-1}}}={\rm e}^{-z}.&(2.9)\cr}
$$Equation (2.9) is the starting point of Tsallis statistics [9]. The left side
in (2.9) is the $q$-exponential function of Tsallis, namely,

$$\eqalignno{[1+(q-1)z]^{-{{1}\over{q-1}}}&=\exp_q(-z)\cr
&={\rm e}^{-z}\hbox{  when  }q\rightarrow 1.&(2.10)\cr}
$$We consider a more general form of (2.9) given by

$$\eqalignno{\lim_{q\rightarrow
1}c|x|^{\gamma}{_1F_0}({{\eta}\over{q-1}};~~;-a(q-1)|x|^{\delta})&=\lim_{q\rightarrow
1}c|x|^{\gamma}[1+a(q-1)|x|^{\delta}]^{-{{\eta}\over{q-1}}}\cr
&=c|x|^{\gamma}{\rm e}^{-a\eta |x|^{\delta}},
a>0,\eta>0,-\infty<x<\infty.&(2.11)\cr}
$$Thus, we obtain Mathai's pathway model [6], namely,

$$f(x)=c|x|^{\gamma}[1+a(q-1)|x|^{\delta}]^{-{{\eta}\over{q-1}}},a>0,\eta>0\eqno(2.12)
$$where $c$ is the normalizing constant. If
$\eta=1,\gamma=0,a=1,\delta=1,x>0$ then (2.12) gives Tsallis
statistics [9]. For
$q>1,a=1,\eta=1,x>0$ in (2.12) we get Beck-Cohen superstatistics [1].

\vskip.2cm It may be observed that (2.7), which is the binomial form
or ${_1F_0}$ going to exponential, is exploited to produce the
pathway model, Tsallis statistics and Beck-Cohen superstatistics whereas (2.6),
which is the Bessel function form going to exponential is not yet
exploited. This produces a rich variety of applicable
functions as will be shown in a forthcoming paper.

\vskip.3cm \noindent {\bf 3.\hskip.3cm Connection of Mittag-Leffler
function to the pathway model}

\vskip.3cm In order to see the connection, let us recall the
Mellin-Barnes representation of the generalized Mittag-Leffler
function.

$$\Gamma(\beta)E_{\alpha,\beta}^{\gamma}(z^{\delta})={{\Gamma(\beta)}\over{\Gamma(\gamma)}}{{1}\over{2\pi
i}}\int_{c-i\infty}^{c+i\infty}{{\Gamma(\gamma-s)\Gamma(s)}\over{\Gamma(\beta-\alpha
s)}}(-z^{\delta})^{-s}{\rm d}s\eqno(3.1)
$$for $\Re(\gamma)>0,i=\sqrt{-1},\Re(\beta)>0$. Let us examine the
situation when $|\beta|\rightarrow\infty$. We will state two basic
results as lemmas.

\vskip.3cm \noindent {\bf Lemma 3.1}\hskip.3cm{\it When
$|\beta|\rightarrow\infty$ and $\alpha >0$ is finite then

$$\lim_{|\beta|\rightarrow\infty}{{\Gamma(\beta)}\over{\Gamma(\beta-\alpha
s)}}[-b(z\beta^{{\alpha}\over{\delta}})^{\delta}]^{-s}=(-bz^{\delta})^{-s}.
\eqno(3.2)
$$}

\vskip.3cm \noindent{\bf Proof.}\hskip.3cm The Stirling's formula is
given by

$$\Gamma(z+a)\approx \sqrt{2\pi}z^{z+a-{1\over2}}{\rm e}^{-z},\hbox{
 for  }|z|\rightarrow\infty\eqno(3.3)
 $$and $a$ is a bounded quantity. Now, applying Stirling's formula
 we have
 $$\eqalignno{{{\Gamma(\beta)}\over{\Gamma(\beta-\alpha
 s)}}[-b(x\beta^{{{\alpha}\over{\delta}}})^{\delta}]^{-s}&\approx
 {{\sqrt{2\pi}\beta^{\beta-{1\over2}}{\rm e}^{-\beta}}\over{\sqrt{2\pi}\beta^{\beta-{1\over2}-\alpha
 s}{\rm e}^{-\beta}}}[-b(x\beta^{{\alpha}\over{\delta}})^{\delta}]^{-s}\cr
 &=\beta^{\alpha s}(-bz^{\delta})^{-s}\beta^{-\alpha
 s}=(-bz^{\delta})^{-s}.&(3.4)\cr}
 $$

 \vskip.3cm
 \noindent
 {\bf Lemma 3.2.}\hskip.3cm{\it For $\Re(\gamma)>0, \Re(\beta)>0$,

 $$\eqalignno{\lim_{|\beta|\rightarrow\infty}\Gamma(\beta)E_{\alpha,\beta}^{\gamma}(-bz^{\delta})&={{1}\over{\Gamma(\gamma)}}{{1}\over{2\pi
 i}}\int_{c-i\infty}^{c+i\infty}\Gamma(\gamma-s)\Gamma(s)[bz^{\delta}]^{-s}{\rm
 d}s&(3.5)\cr
 &=\sum_{k=0}^{\infty}{{(\gamma)_k}\over{k!}}(-bz^{\delta})^k=[1+bz^{\delta}]^{-\gamma}\hbox{
 for  }|bz^{\delta}|<1.&(3.6)\cr}
 $$}

 \vskip.2cm
 \noindent
 {\bf Proof.}\hskip.3cm After applying Lemma 3.1 we can evaluate the
 contour integral in (3.5) by using the residue theorem at the poles
 of $\Gamma(s)$ to obtain the series form in (3.6). For
 $|bz^{\delta}|>1$ we obtain the series form as analytic
 continuation or by evaluating the integral in (3.5) as the sum of
 the residues at the poles of $\Gamma(\gamma-s)$. This will be equal
 to the following:

 $$\lim_{|\beta|\rightarrow\infty}\Gamma(\beta)E_{\alpha,\beta}^{\gamma}(-bz^{\delta})=(bz^{\delta})^{-\gamma}[1+(bz^{\delta})^{-1}]^{-\gamma}\eqno(3.7)
 $$for $|bz^{\delta}|>1$ which will be of the same form as in (3.6).
 Replacing $b$ by $a(q-1)$ and $\gamma$ by ${{\eta}\over{q-1}}$ we
 have the following:

 $$\eqalignno{\lim_{|\beta|\rightarrow\infty}c|z|^{\gamma}\Gamma(\beta)&E_{\alpha,\beta}^{\eta/(q-1)}(-a(q-1)|z|^{\delta})\cr
 &=c|z|^{\gamma}[1+a(q-1)|z|^{\delta}]^{-{{\eta}\over{q-1}}}.&(3.8)\cr}
 $$Observe that (3.8) is nothing but Mathai's pathway model [6] for
 $a>0,\delta>0, \eta>0$ with $c$ being the normalizing constant, for
 both $q>1$ and $q<1$. Then when $q\rightarrow 1$ (3.8) will reduce
 to the exponential form. That is,

 $$\lim_{q\rightarrow
 1}\lim_{|\beta|\rightarrow\infty}c|z|^{\gamma}\Gamma(\beta)E_{\alpha,\beta}^{\eta/(q-1)}[-a(q-1)|z|^{\delta}]
 =c|z|^{\gamma}{\rm e}^{-a\eta |z|^{\delta}}.\eqno(3.9)
 $$Note from (3.8) that when $q>1$ the functional form in (3.8)
 remains the same whether $\delta>0$ or $\delta<0$ with
 $-\infty<z<\infty$. When $q<1$ the support in (3.8) will be
 different for $\delta>0$ and $\delta <0$ if $f(x)$ is to remain as
 a statistical density.

 \vskip.2cm
 For $\gamma=0, a=1,\delta=1,\eta=1,z>0$ we have Tsallis statistics [9]
 coming from (3.8) for both the cases $q>1$ and $q<1$. For $q>1,
 a=1, \delta =1,\eta=1,z>0$ we have Beck-Cohen superstatistics [1] coming from
 (3.8). Observe that the limiting process from (3.5) to (3.6) holds
 for $\gamma=1$ or $\gamma=1$ and $\alpha=1$ also. Thus the special
 cases of Mittag-Leffler function are also covered.

 \vskip.2cm
 Thus, through $\beta$ a pathway is created to go from a generalized
 Mittag-Leffler function to Mathai's pathway model and then to
 Tsallis statistics and Beck-Cohen superstatistics. If $\beta$ is real then as
 $\beta$ becomes larger and larger then

 $$c|z|^{\gamma}\Gamma(\beta)E_{\alpha,\beta}^{\eta/(q-1)}(-a(q-1)|z|^{\delta})
 $$goes closer and closer to the pathway model in (3.8). In other
 words, a pathway is created through $\beta$ to go from a
 Mittag-Leffler function to the pathway model to Tsallis statistics
 and Beck-Cohen superstatistics. Thus for large real value of $\beta$ or for
 large value of $|\beta|$,

 $$c|z|^{\gamma}\Gamma(\beta)E_{\alpha,\beta}^{\eta/(q-1)}[-a(q-1)|z|^{\delta}]\approx
 c|z|^{\gamma}[1+a(q-1)|z|^{\delta}]^{-{{\eta}\over{q-1}}}.\eqno(3.10)
 $$

 \vskip.3cm
 \noindent
 {\bf 4.\hskip.3cm Connections through the H-function}

 \vskip.3cm Recalling the generalized Mittag-Leffler function from
 (3.1) and representing it in terms of a H-function we have the
 following:

 $$\eqalignno{\Gamma(\beta)E_{\alpha,\beta}^{\gamma}(z\beta^{{{\alpha}\over{\delta}}})^{\delta}&={{\Gamma(\beta)}\over{\Gamma(\gamma)}}{{1}\over{2\pi
 i}}\int_{c-i\infty}^{c+i\infty}{{\Gamma(s)\Gamma(\gamma-s)}\over{\Gamma(\beta-\alpha
 s)}}[-(z\beta^{{\alpha}\over{\delta}})^{\delta}]^{-s}{\rm
 d}s&(4.1)\cr
 &={{\Gamma(\beta)}\over{\Gamma(\gamma)}}H_{1,2}^{1,1}\left[-(z\beta^{{\alpha}\over{\delta}})^{\delta}\bigg\vert_{(0,1),
 (1-\beta,\alpha)}^{(1-\gamma,1)}\right].&(4.2)\cr}
 $$From the limiting process discussed in (3.2) we have the
 following result:

 \vskip.3cm
 \noindent
 {\bf Lemma 4.1}\hskip.3cm{\it For $\Re(\beta)>0,\Re(\gamma)>0$,

 $$\eqalignno{\lim_{|\beta|\rightarrow\infty}&{{\Gamma(\beta)}\over{\Gamma(\gamma)}}H_{1,2}^{1,1}
 \left[-(z\beta^{{\alpha}\over{\delta}})^{\delta}\big\vert_{(0,1),(1-\beta,\alpha)}^{(1-\gamma,1)}\right]\cr
 &={{1}\over{\Gamma(\gamma)}}H_{1,1}^{1,1}\left[-z^{\delta}\big\vert_{(0,1)}^{(1-\gamma,1)}\right]&(4.3)\cr
 &={{1}\over{\Gamma(\gamma)}}{{1}\over{2\pi
 i}}\int_{c-i\infty}^{c+i\infty}\Gamma(s)\Gamma(\gamma-s)(-z^{\delta})^{-s}{\rm
 d}s&(4.4)\cr
 &=[1-z^{\delta}]^{-\gamma}.&(4.5)\cr}
 $$}Therefore we have the following theorem:

 \vskip.3cm
 \noindent
 {\bf Theorem 4.1}\hskip.3cm {\it For
 $\Re(\beta)>0,\Re(\gamma)>0,x>0,a>0,q>1,c>0$
 $$\eqalignno{\lim_{|\beta|\rightarrow\infty}c&{{\gamma(\beta)}\over{\Gamma\left({{\eta}\over{q-1}}\right)}}
 x^{\gamma}E_{\alpha,\beta}^{\eta/(q-1)}[-a(q-1)\beta^{{\alpha}\over{\delta}}x]^{\delta}\cr
 &=cx^{\gamma}[1+a(q-1)x^{\delta}]^{-{{\eta}\over{q-1}}}.&(4.6)\cr}
 $$}Observe that the right side in (4.6) is the pathway model [6]
 for $x>0$ from where one has Tsallis statistics, Beck-Cohen superstatistics and power
 law where the constant $c$ can act as the normalizing constant to
 create a statistical density in the right side of (4.6).

 \vskip.2cm We will write the right side in (4.6) as a H-function
 and then establish a connection between Mittag-Leffler function
 and the pathway model through the H-function. To this end, the
 practical procedure is to look at the Mellin transform of the right
 side of (4.6) and then write it as a H-function by taking the
 inverse Mellin transform. The Mellin transform of the right side of
 (4.6), denoted by $M_f(s)$, is given by the following:

 $$\eqalignno{M_f(s)&=\int_0^{\infty}c~x^{\gamma+s-1}[1+a(q-1)x^{\delta}]^{-{{\eta}\over{q-1}}}{\rm
 d}x,a>0,q>1,\eta>0,\delta>0\cr
 &={{c}\over{\delta}}{{\Gamma\left({{\gamma+s}\over{\delta}}\right)}\over{[a(q-1)]^{{\gamma+s}\over{\delta}}}}
 {{\Gamma\left({{\eta}\over{q-1}}-{{\gamma+s)}\over{\delta}}\right)}\over{\Gamma\left({{\eta}\over{q-1}}\right)}}&(4.7)\cr}
 $$for
 $\Re\left({{\gamma+s)}\over{\delta}}\right)>0,\Re\left({{\eta}\over{q-1}}-{{\gamma+s)}\over{\delta}}\right)>0.$
 Note that if $c$ is the normalizing constant for the density $f(x)$
 then by putting $s=1$ the right side of (4.7) must be $1$.
 Therefore,

 $$M_f(s)={{[a(q-1)]^{{1}\over{\delta}}}\over{\Gamma\left({{\gamma+1}\over{\delta}}\right)
 \Gamma\left({{\eta}\over{q-1}}-{{\gamma+1}\over{\delta}}\right)}}\Gamma\left({{\gamma+s}\over{\delta}}\right)
 \Gamma\left({{\eta}\over{q-1}}-{{\gamma+s}\over{\delta}}\right)[a(q-1)]^{-{{s}\over{\delta}}}.\eqno(4.9)
 $$Hence the right side of (4.6) is available as the inverse Mellin
 transform of (4.9). That is,

 $$\eqalignno{f(x)&=cx^{\gamma}[1+a(q-1)x^{\delta}]^{-{{\eta}\over{q-1}}},&(4.10)\cr
 &\hbox{for  } a>0,\eta>0,q>1,\delta>0,x>0\cr
 &={{[a(q-1)]^{{1}\over{\delta}}}\over{\Gamma\left({{\gamma+1}\over{\delta}}\right)\Gamma\left({{\eta}\over{q-1}}-{{\gamma+1}\over{\delta}}\right)}}\cr
 &\times {{1}\over{2\pi
 i}}\int_{c-i\infty}^{c+i\infty}\Gamma\left({{\gamma+s}\over{\delta}}\right)\Gamma\left({{\eta}\over{q-1}}
 -{{\gamma+s}\over{\delta}}\right)[x(a(q-1))^{{1}\over{\delta}}]^{-s}{\rm
 d}s&(4.11)\cr
 &={{[a(q-1)]^{{1}\over{\delta}}}\over{\Gamma\left({{\gamma+1}\over{\delta}}\right)\Gamma\left({{\eta}\over{q-1}}-{{\gamma+1}\over{\delta}}\right)}}
 H_{1,1}^{1,1}\left[x(a(q-1))^{{1}\over{\delta}}\bigg\vert_{({{\gamma}\over{\delta}},{{1}\over{\delta}})}^{(1-{{\eta}\over{q-1}}
 +{{\gamma}\over{\delta}},{{1}\over{\delta}})}\right]&(4.12)\cr
 &=\lim_{|\beta|\rightarrow\infty}c{{\Gamma(\beta)}\over{\Gamma\left({{\eta}\over{q-1}}\right)}}x^{\gamma}
 E_{\alpha,\beta}^{\eta/(q-1)}[-x(a(q-1))^{{1}\over{\delta}}\beta^{{\alpha}\over{\delta}}]^{\delta}&(4.13)\cr
 \noalign{\hbox{where}}
c&={{\delta[a(q-1)]^{{\gamma+1}\over{\delta}}\Gamma\left({{\eta}\over{q-1}}\right)}\over{\Gamma\left({{\gamma+1}\over{\delta}}\right)
 \Gamma\left({{\eta}\over{q-1}}-{{\gamma+1}\over{\delta}}\right)}}&(4.14)\cr}
 $$for
 $q>1,\delta>0,\eta>0,a>0,\Re(\gamma+1)>0,\Re\left({{\eta}\over{q-1}}-{{\gamma+1}\over{\delta}}\right)>0.$

 \vskip.3cm
 \noindent
 {\bf Remark 4.1.}\hskip.3cm From (4.13) and (4.10) it can be noted
 that as the parameter $\beta$ becomes larger and larger the
 Mittag-Leffler function in (4.13) comes closer and closer to the
 pathway model and eventually in the limiting situation both become
 identical. In a physical situation, if (4.10) represents the stable
 situation then the unstable neighborhoods are given by (4.13). When
 $q\rightarrow 1$ then (4.10) goes to the exponential form. Thus if
 the exponential form is the stable
 situation then the pathway model of (4.1) itself models the
 unstable neighborhoods. This unstable neighborhood is farther
 extended by the Mittag-Leffler form in (4.13).

 \vskip.3cm
 \noindent
 {\bf Remark 4.2.}\hskip.3cm Observe that the functional form in
 (4.10) remains the same whether $q>1$ or $q<1$. But for $q<1$ or
 when $q\rightarrow 1$ the normalizing constant $c$ will be
 different.

 \vskip.3cm
 \noindent
 {\bf Remark 4.3.}\hskip.3cm When dealing with problems such as
 reaction-diffusion situations or a general production-destruction model one
 goes to fractional differential equations to get a better picture
 of the solution. Then we usually end up in Mittag-Leffler functions
 and their generalizations into Wright's function or Fox function. Comparison of
 (4.13) and (4.10) reveals that as $\beta$ gets larger and larger
 the effect of fractional derivative becomes less and less and
 finally when $|\beta|\rightarrow\infty$ the effect of taking
 fractional derivatives, instead of total derivatives, gets nullified.
 
\vskip.3cm \noindent\centerline{\bf References}

\vskip.3cm \noindent [1] Beck, C. Superstatistics: Theoretical Concepts and Physical Applications, in R. Klages, G. Radons, and I.M. Sokolpv (Eds.): Anomalous Transport: Foundations and Applications, Wiley-VCH, Weinheim, 2008, pp. 433-457.

\vskip.3cm \noindent [2] Jorgenson, J. and Lang, S. The ubiquitous heat kernel, in B. Engquist and W. Schmid (Eds.): Mathematics Unlimited - 2001 and Beyond, Springer, Berlin and Heidelberg, 2001, pp. 655-683.

\vskip.3cm \noindent [3] Kilbas, A.A., Saxena, R.K., Saigo, M. and Trujillo, J.J. Generalized Wright function as the H-function, in A.A. Klages and S.V. Rogosin (Eds.): Analytic Methods of Analysis and Differential Equations, Cambridge Scientific Publishers Ltd, Cambridge, 2006, pp. 117-134.

\vskip.3cm \noindent [4] Haubold, H.J. and Mathai, A.M. The fractional kinetic equations and thermonuclear functions, Astrophysics and Space Science, 273(2000), 53-63. Saxena, R.K., Mathai, A.M. and Haubold, H.J. On fractional kinetic equations, Astrophysics and Space Science, 282(2000), 281-287. Saxena, R.K., Mathai, A.M. and Haubold, H.J. On generalized fractional kinetic equations, Physica A, 344(2004), 657-664. Saxena, R.K., Mathai, A.M. and Haubold,
H.J. Unified fractional kinetic equation and a fractional diffusion equation, Astrophysics and Space Science, 290(2004), 299-310.

\vskip.3cm \noindent [5] Mathai, A.M. and Haubold H.J. Special Functions for Applied Scientists, Springer, New York, 2008.

\vskip.3cm \noindent [6] Mathai, A.M. A pathway to matrix-variate gamma and Gaussian densities, Linear Algebra and Its Applications, 396(2005),317-328. Mathai,A.M. and Haubold, H.J. On generalized entropy measures and pathways. Physica A, 385(2007), 493-500. Mathai, A.M. and Haubold, H.J. Pathway model, superstatistics, Tsallis statistics and a generalized measure of entropy, Physica A, 375(2007), 110-122. Mathai, A.M. and Haubold, H.J. On generalized distributions and pathways, Physics Letters A, 372 (2008), 2109-2113. Mathai, A.M. and Haubold, H.J. Pathway parameter and thermonuclear functions, Physica A, 387 (2008), 2462-2470.

\vskip.3cm \noindent [7] Mathai, A.M., Saxena, R.K. and Haubold, H.J. The H-Function: Theory and Applications, Springer, New York, 2009.

\vskip.3cm \noindent [8] Schwartz, C. Nonlinear operators and their propagators, Journal of Mathematical Physics, 38(1997), 484-500. Schwartz, C. Nonlinear operators II, Journal of Mathematical Physics, 38(1997) 3841-3862.

\vskip.3cm \noindent [9] Tsallis, C. Introduction to Nonextensive Statistical Mechanics: Approaching a Complex World, Springer, New York, 2009.

\vskip.3cm

\bye